\documentclass[conference]{IEEEtran}

\IEEEoverridecommandlockouts
\usepackage{verbatim}
\usepackage{longtable}
\DeclareMathAlphabet{\mathpzc}{OT1}{pzc}{m}{it}
\usepackage{amsmath}
\usepackage{amsfonts}
\usepackage{amssymb}
\usepackage{multirow}
\usepackage{amsmath,amssymb,amsfonts}
\usepackage{fancyhdr}
\usepackage[ruled,vlined,linesnumbered]{algorithm2e}
\usepackage[
top    = 0.7in,
bottom = 1in,
left   = 0.65in,
right  = 0.59in]{geometry}
\usepackage{graphicx}
\usepackage{textcomp}
\usepackage{xcolor}
\usepackage{subcaption}
\usepackage{algpseudocode}
\usepackage{enumitem}
\usepackage{float}
\restylefloat{figure}
\usepackage{graphicx}
\usepackage{url}
\usepackage[normalem]{ulem}
\useunder{\uline}{\ul}{}

\begin{document}

\title{\texttt{\textbf{SWORD:}} A \textbf{S}ecure Lo\textbf{W}-Latency \textbf{O}ffline-First Authentication and Data Sharing Scheme for \textbf{R}esource Constrained \textbf{D}istributed Networks}

\author{\IEEEauthorblockN{Faisal Haque Bappy$^{1}$, Tahrim Hossain$^{2}$, Raiful Hasan$^{3}$, Kamrul Hasan$^{4}$,\\Mohamed Younis$^{5}$, Tariqul Islam$^{6}$}
\IEEEauthorblockA{
$ ^{1, 2, 5, 6}$ University of Maryland Baltimore County, MD, USA\\
$ ^{3}$ Kent State University, OH, USA; 
$ ^{4}$ Tennessee State University, TN, USA\\
Email: \{fbappy1@umbc, m482@umbc, rhasan7@kent, mhasan1@tnstate, younis@umbc, mtislam@umbc\}.edu}}

\maketitle

\thispagestyle{fancy}
\chead{This work has been accepted at the 2026 IEEE International Conference on Communications (ICC)}
\cfoot{}

\pagestyle{empty}

\begin{abstract}
While many resource-constrained networks, such as Internet of Things (IoT) and Internet of Vehicles (IoV), are inherently distributed, the majority still rely on central servers for fast authentication and data sharing. Blockchain-based solutions offer decentralized alternatives but often struggle to meet the stringent latency requirements of real-time applications. Even with the rollout of 5G, network latency between servers and peers remains a significant challenge. To address this, we introduce \texttt{\textbf{SWORD}}, a novel offline-first authentication and data-sharing scheme designed specifically for resource-constrained networks. \texttt{\textbf{SWORD}} utilizes a proximity-based clustering approach to enable offline authentication and data sharing, ensuring low-latency, secure operations even in intermittently connected scenarios. Our experimental results show that \texttt{\textbf{SWORD}} outperforms traditional blockchain-based solutions while offering similar resource efficiency and authentication latency to central-server-based solutions. Additionally, we provide a comprehensive security analysis, demonstrating that \texttt{\textbf{SWORD}} is resilient against spoofing, impersonation, replay, and man-in-the-middle attacks.
\end{abstract}


\section{Introduction}
\label{sec:intro}
The rise of resource-constrained devices has accelerated the development of IoT and IoV systems, where devices with limited processing power and energy communicate to perform essential tasks. Although these systems are distributed in nature, they often rely on centralized servers for authentication and data exchange to ensure low-latency responses. This centralized approach, however, introduces several issues such as single points of failure, vulnerability to DDoS attacks, and privacy risks due to centralized data storage \cite{butun2019security}.

While traditional certificate-based schemes \cite{mundhe2020efficient,qiao2023anonymous} and decentralized databases \cite{tiennoy2018using} have been explored for secure communication in IoT and VANET systems, they suffer from critical limitations in dynamic and disconnected environments. Certificate-based models require constant interaction with trusted authorities for issuance and revocation, which is infeasible in regions lacking infrastructure or in scenarios where connectivity is intermittent. Decentralized databases, while useful for distributing data, typically lack built-in tamper resistance, immutability, and transparent audit trails, properties essential for trustless authentication and secure data sharing in adversarial environments.

In contrast, blockchain provides a tamper-evident ledger for authentication records and credential management, enabling decentralized consensus without centralized authentication servers. However, traditional blockchain systems often fail to meet the low-latency demands of real-time IoT and IoV applications, with authentication times ranging from seconds to minutes, unsuitable for safety-critical tasks like vehicle-to-vehicle (V2V) communication \cite{ayed2023blockchain}. Even with 5G, network latency remains a bottleneck in large-scale deployments \cite{schulz2017latency}.

To address performance and security, recent efforts have introduced lightweight consensus mechanisms \cite{li2018blockchain}, smart contract automation \cite{gong2021blockchain, goyat2020blockchain}, and dynamic authentication schemes \cite{xu2020mutual, studer2007efficient}. While effective, these approaches depend on continuous connectivity, limiting their applicability in disconnected environments. Blockchain-based IoT authentication \cite{lau2018blockchain} and VANET security frameworks still prioritize online workflows, overlooking challenges such as intermittent connectivity, node isolation, and sparse topologies. This highlights the urgent need for offline-first solutions, as emphasized in recent work on efficient offline authentication protocols \cite{hou2023efficient}.

To address these challenges, we introduce \texttt{\textbf{SWORD}}, a secure and low-latency offline-first authentication and data sharing scheme designed for resource-constrained distributed environments. \texttt{\textbf{SWORD}} employs blockchain technology to provide tamper-evident audit trails for authentication events, enable decentralized consensus through offline cluster verification, and ensure eventual consistency when clusters resynchronize with the main network. Unlike traditional blockchain applications that use the ledger for data storage, \texttt{\textbf{SWORD}} leverages blockchain exclusively for authentication credential management and maintaining immutable authentication records. \texttt{\textbf{SWORD}} features a novel proximity-based clustering mechanism that supports decentralized, tamper-resistant authentication even in the absence of continuous connectivity. Unlike traditional online-dependent approaches, \texttt{\textbf{SWORD}} adapts dynamically to network conditions, enabling robust operation in both connected and disconnected scenarios. Our framework represents a significant step toward scalable, offline-first security solutions that balance performance, availability, and trust in resource-limited systems. The main contributions of this paper are as follows:

\begin{itemize}
    \item \textbf{Offline-First Blockchain Framework:} We propose \texttt{\textbf{SWORD}}, a secure and low-latency authentication and data sharing scheme tailored for resource-constrained distributed networks. \texttt{\textbf{SWORD}} is designed to operate effectively in both connected and disconnected settings by prioritizing offline-first authentication workflows.
    
    \item \textbf{Proximity-Based Clustering for Resilience and Efficiency:} We introduce a novel proximity-aware clustering mechanism that enables secure local interactions among nearby nodes, reduces reliance on global connectivity, and ensures scalable authentication and data exchange in sparse or partitioned topologies.
    
    \item \textbf{Comprehensive Evaluation and Security Analysis:} We validate the effectiveness of \texttt{\textbf{SWORD}} through extensive experimental evaluation, demonstrating low-latency performance and resource efficiency. We also provide a formal security analysis, showing resilience against spoofing, impersonation, and man-in-the-middle attacks.
\end{itemize}

The rest of the paper is organized as follows:  Section \ref{sec:lit} reviews the existing literature. Section \ref{sec:arch} describes the architecture of our proposed scheme, \texttt{\textbf{SWORD}} in detail. The implementation and performance analysis are discussed in Section \ref{sec:perf}, followed by a security analysis in Section \ref{sec:sec}. Finally, we conclude the paper in Section \ref{sec:conc}.

\section{Related Works}
\label{sec:lit}
Centralized authentication frameworks have long dominated IoT and IoV systems due to their rapid response times, with protocols like OAuth 2.0 achieving authentication latencies of 50-100 ms under optimal conditions \cite{fett2016comprehensive}. However, these systems suffer from critical vulnerabilities, including single points of failure and susceptibility to DDoS attacks \cite{drame2021centralized, butun2019security}. While certificate-based approaches using PKI offer reduced server dependence with authentication times of 150-300 ms \cite{farooq2019certificate}, they face challenges in certificate management and computational overhead on resource-constrained devices \cite{hasrouny2017vanet}. Traditional blockchain solutions promise decentralization but introduce prohibitive 3-10 second consensus delays \cite{lau2018blockchain} that render them impractical for real-time IoV applications \cite{alqahtani2021bottlenecks}.

Although several researchers have explored blockchain-based authentication solutions, significant gaps remain. Li et al. \cite{li2018blockchain} proposed a lightweight consensus mechanism for resource-constrained IoT devices, while Gong et al. \cite{gong2021blockchain} and Goyat et al. \cite{goyat2020blockchain} introduced smart contracts for access control and privacy-preserving storage; however, all assume continuous network availability, overlooking offline authentication critical for intermittent connectivity scenarios. In vehicular networks, existing work addresses dynamic authentication \cite{xu2020mutual}, convoy scenarios \cite{studer2007efficient}, and DoS mitigation \cite{he2012mitigating}, but fails to address efficient authentication in resource-constrained, intermittently connected environments.

Clustered blockchain architectures \cite{ramamoorthi2023energy, ayed2023blockchain} partition networks into smaller groups that process transactions in parallel, reducing congestion and improving throughput. Recent W3C standards for Decentralized Identifiers (DIDs) \cite{sporny2022did} provide portable, cryptographically verifiable identity solutions. While \texttt{\textbf{SWORD}} incorporates DIDs, it differs from VC-based frameworks through proximity clusters for distributed verification, a Temporary Authentication Ledger for eventual consistency, and threshold-based consensus optimized for IoT constraints \cite{alamer2022efficient}. Collectively, existing solutions reveal a critical gap: the need for a unified framework supporting secure, decentralized, offline authentication across diverse and intermittently connected environments.

\begin{figure*}[!htbp]
\centering
\includegraphics[width=\textwidth]{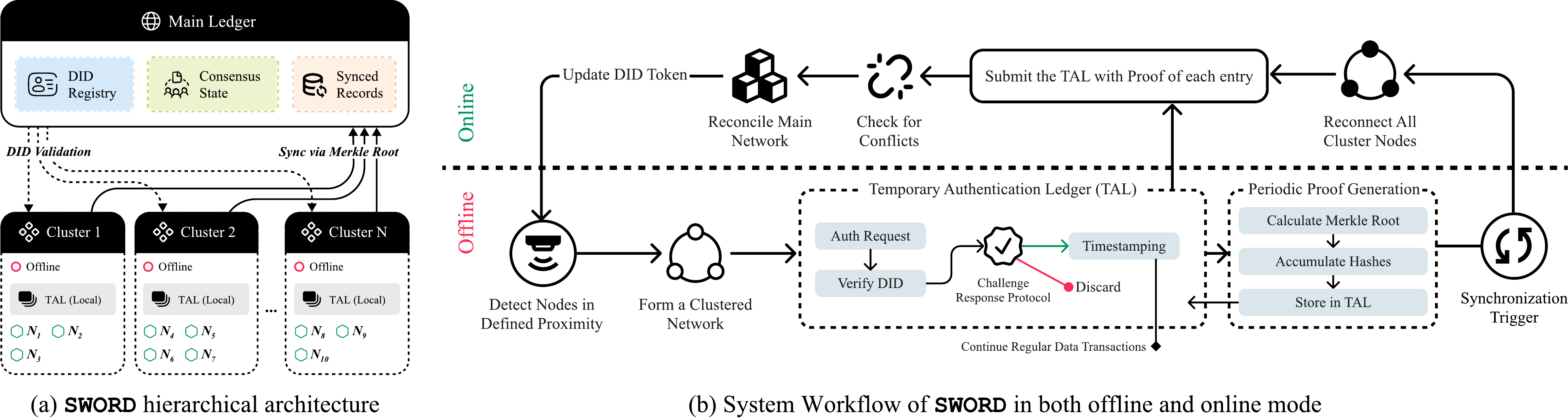}
        \caption{\texttt{\textbf{SWORD}} architecture and workflow showing hierarchical cluster and authentication flow with periodic synchronization.}
\label{fig:arch}
\end{figure*}

\section{Architectural Overview} 
\label{sec:arch}

This section presents the design of \texttt{\textbf{SWORD}}, a decentralized, low-latency authentication and data-sharing scheme for resource-constrained networked systems. Fig.~\ref{fig:arch} illustrates the system's hierarchical architecture and operational workflow. As shown in Fig.~\ref{fig:arch}a, \texttt{\textbf{SWORD}} employs a two-tier structure where proximity-based offline clusters operate semi-autonomously under a global blockchain network. The system workflow (Fig.~\ref{fig:arch}b) demonstrates how five key components enable secure offline authentication: proximity-based clustering for autonomous local groups, Temporary Authentication Ledgers for minimal-storage event recording, challenge-response protocols with threshold-based voting, Merkle-proof\cite{merkle1989certified} synchronization for efficient global blockchain merging, and a permissioned trust model ensuring node honesty. This layered approach solves offline authentication without sacrificing blockchain security guarantees.

\subsection{Proximity-based Clustering Mechanism}
Like other decentralized ledger systems, \texttt{\textbf{SWORD}} maintains a global blockchain network but introduces proximity-based clustering. We adopt this approach because geographically close nodes are more likely to remain connected during network partitions, reducing synchronization conflicts and improving offline authentication success rates compared to random or load-based clustering. When nodes detect nearby peers, they form autonomous offline clusters that maintain localized blockchain ledgers via short-range wireless communication. Each cluster's TAL uses lightweight storage: authentication records (device ID, timestamp, hash) require $\sim$256 bytes, accumulating only $\sim$250 KB for 1,000 daily authentications. Merkle tree computation for 1,000 entries requires 10 hash operations ($\log_2 1000 \approx 10$), completing in $<$50ms on ARM processors. Synchronized clusters automatically prune confirmed entries, maintaining local storage $<$1 MB.

\subsection{Cluster Formation and Membership Management}
\label{subsec:cluster_formation}
Nodes periodically broadcast beacon messages containing their DID and blockchain state hash via Wi-Fi Direct or Bluetooth LE (range: 10-100m). Receiving nodes verify the broadcaster's DID against their local blockchain copy; only validated nodes proceed. When $\geq 3$ mutually reachable nodes detect each other, they perform lightweight consensus to establish the cluster roster, creating a timestamped, signed membership certificate. We require a minimum of 3 nodes to prevent trivial two-node clusters vulnerable to deadlock, while majority voting ($\geq \frac{2}{3}$) for new members ensures Byzantine fault tolerance without unanimous consensus overhead that becomes impractical as clusters grow. Alternative approaches like RAFT or Paxos were ruled out due to their complexity and latency unsuitable for resource-constrained devices\cite{alqahtani2021bottlenecks}. New nodes entering proximity require $\geq \frac{2}{3}$ votes from existing members after DID validation. Departing nodes (detected via 3 missed 30-second heartbeats) are marked ``departed'' in the TAL; their prior contributions remain valid, but they cannot participate in new authentications. The authentication threshold adapts dynamically: $t = \lceil 0.6 \times n_{\text{current}} \rceil$. If $n < 3$, the cluster enters ``degraded mode'' requiring mandatory online verification. All join/leave events are timestamped in the TAL, and membership snapshots are hashed into authentication records for tamper detection. During synchronization, the main network validates membership changes and checks for anomalies indicating potential attacks.

\subsection{Temporary Authentication Ledger}
Each offline cluster maintains a Temporary Authentication Ledger (TAL), a lightweight repository that temporarily stores authentication events locally to enable rapid authentication without network latency or synchronization overhead. The TAL records device IDs, timestamps, and corresponding hashes for future synchronization with the main network. Within offline clusters, \texttt{\textbf{SWORD}} supports \textbf{decentralized identifiers (DIDs)} for device verification, with all DIDs securely stored in the global blockchain. Offline clusters rely on pre-synchronized blocks in local storage to verify credentials. When a device attempts authentication in an offline environment, its request is processed locally and stored in the TAL, ensuring rapid authentication without real-time validation from the main network.

\IncMargin{1em}
\begin{algorithm}[]
\caption{Challenge-Response Protocol}
\label{alg:challenge}
\KwIn{Device $D$, Cluster $C = \{N_1, N_2, \dots, N_n\}$, Public Key $PK_D$, Private Key $SK_D$, Threshold $t$}
\KwOut{Authentication Result}
\SetKwFunction{GenRandChallenge}{GenRandChallenge}
\SetKwFunction{Sign}{Sign}
\SetKwFunction{Send}{Send}
\SetKwFunction{Verify}{Verify}
\SetKwFunction{LogEvent}{LogEvent}
Challenges $\leftarrow \{\}$; Responses $\leftarrow \{\}$; ValidCount $\leftarrow 0$\\
\ForEach{$N_i \in C$}{
    $R_i \leftarrow \GenRandChallenge(\text{mfa}(N_i))$; Challenges $\leftarrow$ Challenges $\cup \{R_i\}$; $\Send(R_i\rightarrow D)$
}
\ForEach{$R_i \in \text{Challenges}$}{
    $S_i \leftarrow \Sign(SK_D, R_i)$; $\Send(S_i \rightarrow N_i)$; $Responses \leftarrow Responses \cup \{(R_i, S_i)\}$
}
\ForEach{$(R_i, S_i) \in \text{Responses}$}{
    \If{$\Verify(PK_D, R_i, S_i)$}{ValidCount $\leftarrow$ ValidCount + 1}
}
\eIf{ValidCount $\geq t$}{
    \LogEvent($D, Success$) \tcp{Grant access}
}{
    \LogEvent($D, Failure$) \tcp{Deny access}
}
\end{algorithm}
\DecMargin{1em}

\subsection{Challenge-Response Protocol}
To prevent replay attacks and prove private key possession without exposure, \texttt{\textbf{SWORD}} employs a challenge-response protocol for identity verification (Algorithm \ref{alg:challenge}). Alternative approaches like zero-knowledge proofs\cite{fiege1987zero} were avoided due to computational overhead. Upon authentication initiation, the cluster generates challenges based on the device's DID and MFA credentials using local time-based OTPs with pre-shared secrets (lines 1-3). Each node creates a random challenge for the device to sign and return (lines 4-5). Authentication succeeds if at least threshold $t$ of $n$ node responses verify successfully, ensuring decentralized verification robust against node failures or compromises. Each offline cluster periodically computes a Merkle root of its TAL as a compact proof for global synchronization. Conflicts (e.g., simultaneous authentications across clusters) are resolved using timestamps, prioritizing earlier transactions. While alternative blockchain conflict resolution protocols exist \cite{xuLocking, fabricCRDT}, we chose timestamping for resource efficiency and faster consensus.

\subsection{Periodic Synchronization}
At regular intervals, each offline cluster initiates synchronization with the global network upon connectivity restoration, as shown in Algorithm~\ref{alg:sync}. The process begins by generating a Merkle root of the TAL (line 1) and sending it to the main network for verification (line 2). Upon successful verification, conflicts with the main ledger (ML) are resolved by ordering transactions via timestamps and updating the ML (lines 4-5). A confirmation is then sent back to the cluster. If verification fails, synchronization is rejected and a re-transmission is requested (lines 8-9), ensuring consistency between offline and global ledgers.

\subsection{Trust Model and Assumptions}
\label{subsec:trust_model}

\texttt{\textbf{SWORD}} operates within Hyperledger Fabric's permissioned consortium model, inheriting strong identity guarantees for secure offline operation. Our trust model relies on five key assumptions:

\textbf{Bootstrap Trust:} Nodes register through Fabric's Membership Service Provider (MSP), receiving X.509 certificates binding cryptographic keys to verified organizational identities. Each node's DID anchors to this certificate, establishing an auditable identity chain before any clustering operations.

\textbf{Cluster Formation Trust:} Only nodes with valid DID entries in the synchronized blockchain state can form or join offline clusters, eliminating complex negotiation protocols and preventing compromised or revoked credentials from participating.

\textbf{Byzantine Fault Tolerance:} We assume honest majority ($>50\%$) within offline clusters, reasonable in permissioned IoT/IoV deployments with Fabric's MSP identity verification and organizational trust (e.g., city infrastructure). Threshold-based authentication requires $t > n/2$ valid responses for correctness despite minority failures. We mitigate collusion attacks through: validated DID restrictions, dynamic $t$ adjustment, post-synchronization audits, and geographic/organizational diversity requirements. This assumption may fail in highly adversarial environments.

\IncMargin{1em}
\begin{algorithm}[]
\caption{Sync Cluster with Main Network}
\label{alg:sync}
\KwIn{Temporary Authentication Ledger $TAL$, Main Ledger $ML$}
\KwOut{Synchronized Ledgers}

\SetKwFunction{GenMerkleRoot}{GenMerkleRoot}
\SetKwFunction{VerifyMerkleRoot}{VerifyMerkleRoot}
\SetKwFunction{ResolveConflicts}{ResolveConflicts}
\SetKwFunction{UpdateLedger}{UpdateLedger}
\SetKwFunction{send}{Send}

$MR \leftarrow \GenMerkleRoot(TAL)$ \\
$\send(MR, TAL \rightarrow Main Network)$\\

\If{$\VerifyMerkleRoot(MR, TAL)$}{
    Conflicts $\leftarrow \ResolveConflicts(TAL, ML)$\\
    $ML \leftarrow \UpdateLedger(ML, TAL, Conflicts)$\\
    \texttt{ConfirmCluster}() \\
}
\Else{
    \texttt{RejectSync}()\\
    \texttt{Retransmit}()\\
}

\end{algorithm}
\DecMargin{1em}

\textbf{Permissioned Network Benefits:} Fabric's identity verification prevents Sybil attacks and enables accountability through audit trails, prioritizing security and traceability for enterprise IoT and IoV. Unlike sidechains, our offline cluster mechanism is ephemeral, proximity-based, and optimized for resource-constrained devices.

\textbf{Communication Capabilities:} We assume nodes possess short-range wireless communication capabilities (Wi-Fi, Bluetooth, or similar) for proximity-based cluster formation, which is standard in modern IoT/IoV devices, including Raspberry Pi, ESP32, and vehicle OBUs.

\section{Implementation and Performance Analysis}
\label{sec:perf}
We implemented \texttt{\textbf{SWORD}} on Hyperledger Fabric \cite{fabric} using custom chaincode and private data collections to manage the Temporary Authentication Ledger (TAL) within offline clusters\footnote{The source code of \texttt{\textbf{SWORD}}: \url{https://github.com/SPaDeS-Lab/SWORD}}. Each cluster functions as an independent channel, supporting autonomous operation and periodic synchronization with the main network. The core chaincode extends Fabric's Membership Service Provider with a threshold-based authentication mechanism. Synchronization occurs via inter-channel communication, where clusters share Merkle roots of their TAL. Upon connectivity restoration, the chaincode triggers synchronization, propagating authentication records to the main channel. A timestamp-based conflict resolution protocol, implemented in the main channel, validates Merkle proofs and resolves conflicts before committing to the global state.

\begin{figure*}[!htbp]
\centering
\includegraphics[width=\textwidth]{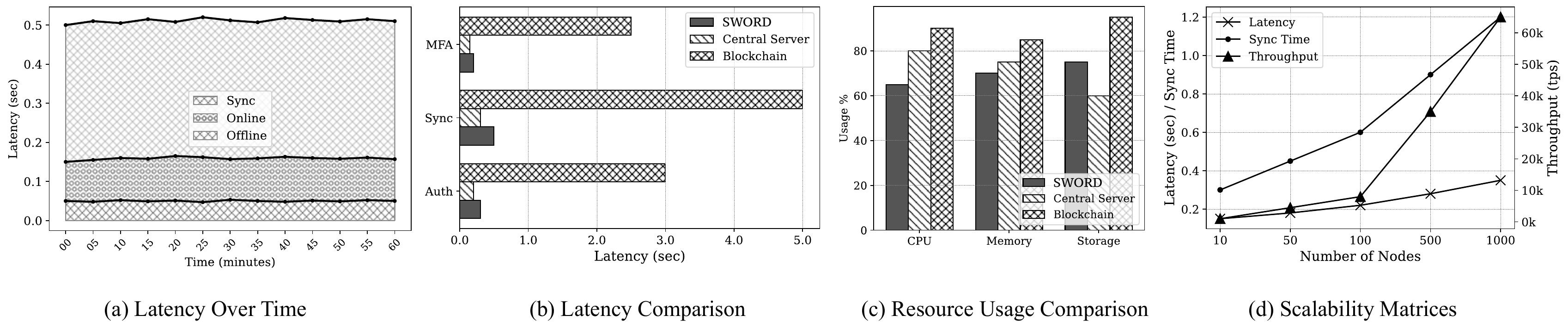}
        \caption{\texttt{\textbf{SWORD}}'s performance over time, its comparison with traditional approaches, and scalability with increasing nodes} \label{fig:perf}
\end{figure*}

\subsection{Experimental Environment}

Our implementation was tested across network configurations scaling from 10 nodes per cluster to 1000 nodes globally, running on edge gateway nodes (Raspberry Pi 4: dual-core ARM, 2GB RAM, 32GB storage, Xubuntu 20.04), representative of IoV on-board units and smart city edge servers, not resource-minimal sensors. Resource-constrained IoT sensors connect through these gateways rather than running full blockchain nodes. The testbed compared three setups: clustered \texttt{\textbf{SWORD}}, traditional blockchain (Hyperledger Fabric), and centralized authentication (OAuth 2.0).

To measure authentication latencies, we generated authentication requests at varying rates (100-1000 requests/minute). For offline scenarios, we isolated clusters from the main network and recorded local authentication latency. Online authentication measured the complete round-trip time with full network connectivity. Synchronization latency was measured by reintroducing offline clusters and timing the complete state synchronization process. Resource utilization (CPU, memory, storage) was collected using Grafana JS API across all nodes. For scalability testing, we incrementally increased network size from 10 to 1000 nodes while maintaining consistent workload, measuring impact on latency, throughput (TPS), and synchronization time.

\subsection{Performance Analysis}
Our performance analysis demonstrated \texttt{\textbf{SWORD}}'s effectiveness across different operational scenarios. Resource utilization metrics (Fig. \ref{fig:perf}c) represent average consumption during mixed workloads: online authentication (40\%), offline authentication (40\%), and synchronization (20\%). Authentication latency measurements revealed consistent patterns: offline authentication achieved 150-200 ms, while synchronization showed 500-600 ms (Fig. \ref{fig:perf}a). Compared to traditional blockchain and centralized servers, our clustered solution showed balanced performance (Fig. \ref{fig:perf}b) and uniquely supported offline authentication, though centralized servers were faster online.

These results generalize across IoT and IoV deployments with similar hardware and workload distributions. Cluster size variations (10-100 nodes) showed less than 15\% deviation in resource patterns. Operational modes exhibited distinct characteristics: offline mode had lower CPU usage and minimal network consumption, online mode required higher bandwidth with moderate CPU usage, and synchronization caused temporary spikes across all metrics.

\texttt{\textbf{SWORD}}'s clustered approach achieved balanced resource consumption across CPU, memory, and storage (Fig. \ref{fig:perf}c). Scalability tests showed efficient handling of increased node counts: latency scaled from 150 ms to 350 ms, throughput improved from 1,000 to 65,000 TPS (Fig. \ref{fig:perf}d), and synchronization time increased linearly, indicating predictable scaling. These results validate \texttt{\textbf{SWORD}}'s viability for resource-constrained IoT and IoV environments.

\section{Security Analysis}
\label{sec:sec}
The security analysis of \texttt{\textbf{SWORD}} examines the framework's approach to ensuring data integrity and secure authentication. \texttt{\textbf{SWORD}} employs a combination of cryptographic techniques and access control measures to defend against potential threats. This analysis offers an overview of these security mechanisms.

\subsection{Credential Validation}

Credential validation safeguards access by preventing unauthorized infiltration. Our framework enforces conditions to ensure only authenticated participants with valid credentials gain access, as outlined in the following lemma.

\textit{\textbf{Lemma 1: }Let \( D = \{ d_1, d_2, \ldots, d_n \} \) be the set of authenticated devices, with \( N \subset D \) as the offline subset. Each node \( n_i \in N \) retains a synchronized copy of the DID token block \( B_D \), periodically updated from the online blockchain. \( B_D \) contains time-bound session credentials for each device in \( D \). If an unauthorized entity \( x \notin D \) attempts to join the offline network cluster, all nodes in \( N \) will deny \( x \) due to the absence of valid credentials in \( B_D \), enforcing secure credential-based authentication.}

\textbf{Proof: } Each node maintains a list $B_D$ of valid credentials, where each entry contains a device ID and expiration time: $B_D = \{(\text{DID}_i, t_i)\}$. To join the network, a device must have a valid, unexpired credential in $B_D$. Consider an unauthorized device $x \notin D$. Since $x$ is not authorized, its credential is not in the list: $(\text{DID}_x, t_x) \notin B_D$. The verification function returns True only if the credential exists in $B_D$ and has not expired. Since $x$'s credential is not in $B_D$, all nodes reject $x$:  $$V(\text{DID}_x) = \text{False} \quad \forall n_i \in N$$ Therefore, unauthorized devices cannot join the network. $\square$

\texttt{\textbf{SWORD}}'s credential validation also defends against the following attacks.

\textbf{i. Spoofing and Impersonation:} For any entity \( y \) to gain access, it must have a valid entry \( (\text{DID}_y, t_y) \in B_D \) with \( t_y > t_{\text{now}} \), ensuring that \( V(\text{DID}_y) = \text{True} \). In contrast, an outsider \( x \notin D \) lacks a valid credential, meaning all nodes \( n_i \in N \) will deny access, as \( V(\text{DID}_x) = \text{False} \) across the network. This mechanism effectively blocks spoofing and impersonation attempts, allowing only authenticated devices in \( D \) with entries in \( B_D \) to be accepted.

\textbf{ii. Replay Attacks:} Each entry \( (\text{DID}_y, t_y) \in B_D \) is time-bound, requiring \( t_y > t_{\text{now}} \) for validity, so \( V(\text{DID}_y) = \text{True} \) only if \( t_y > t_{\text{now}} \). If a token is reused after expiration, it will be rejected by all \( n_i \in N \), mitigating replay attacks. 
\subsection{Data Integrity Verification}

Data integrity is essential for maintaining trust and consistency within the system. \texttt{\textbf{SWORD}}'s use of a Merkle tree \cite{merkle1989certified} allows any alteration to stored data to be detected, ensuring data integrity and network security, as outlined in the following lemma.

\textit{\textbf{Lemma 2: }Let \( N \) be an offline cluster that periodically organizes its transactions \( \{ T_1, T_2, \ldots, T_k \} \) into a Merkle tree, producing a Merkle root \( R \). Upon reconnection to the online network, \( R \) serves as a commitment to the integrity of \( \{ T_1, T_2, \ldots, T_k \} \). Any transaction \( T_j \) can be subsequently verified against \( R \), ensuring data integrity within \( N \) during offline periods.}

\textbf{Proof:} Each transaction $T_j$ in the offline cluster is hashed to create a leaf node: $H_{0,j} = \texttt{Hash}(T_j)$ for $j = 1, 2, \dots, k$. Parent nodes are formed by hashing concatenated pairs: $H_{i,j} = \texttt{Hash}(H_{i-1,2j-1} \| H_{i-1,2j})$. This continues for $\log_2(n)$ levels until a single Merkle root $R$ remains. If any transaction $T_j$ is modified, its leaf hash changes: $H_{0,j} \to H'_{0,j} \neq H_{0,j}$. This change propagates upward through all parent nodes, ultimately changing the root: $R \to R' \neq R$. Therefore, any modification to a transaction is detected by the changed Merkle root, preserving data integrity. $\square$

\texttt{\textbf{SWORD}}'s Merkle tree structure also ensures resilience against the following attacks.

\textbf{i. Data Tampering:} The Merkle root \( R \) verifies the integrity of the transaction set \( \{ T_1, T_2, \dots, T_k \} \) within \( N \), ensuring data consistency before syncing with the main network. If any transaction \( T_j \) is accidentally or intentionally altered within \( N \), such as \( T_j \to T_j' \), the resulting hash \( H'_{0,j} = \texttt{Hash}(T_j') \) will differ from the original \( H_{0,j} \), leading to a different root \( R' \neq R \). This mismatch between \( R' \) and \( R \) signals unauthorized changes within the cluster, thereby preserving data integrity upon reconnection.

\textbf{ii. Man-in-the-Middle (MITM) Attack:} In an MITM attack, an external adversary intercepts and alters transactions, such as \( T_j \to T_j' \), during transmission from \( N \) to the main network. This modification changes the hash \( H'_{0,j} = \texttt{Hash}(T_j') \), resulting in a modified root \( R' \neq R \). When the network detects this mismatch between \( R' \) and \( R \), it reveals tampering during transmission, thereby exposing the MITM attack.

\section{Conclusion}
\label{sec:conc}
In this paper, we introduced \texttt{\textbf{SWORD}}, an offline-first, clustered blockchain authentication and data-sharing framework for resource-constrained distributed networks. Leveraging proximity-based clustering, \texttt{\textbf{SWORD}} enables secure, low-latency interactions in both connected and disconnected settings. Our Hyperledger Fabric implementation achieves authentication latency and resource efficiency comparable to centralized systems while eliminating single points of failure, and significantly outperforms traditional blockchain solutions in real-time responsiveness. Security analysis demonstrates resilience against spoofing, impersonation, replay, tampering, and man-in-the-middle attacks. Future work includes integrating W3C Verifiable Credentials, exploring advanced conflict resolution mechanisms, and investigating adaptive threshold mechanisms. \texttt{\textbf{SWORD}} provides a practical foundation for offline-capable IoT, IoV, and edge systems.

\bibliographystyle{IEEEtran}
\bibliography{IEEEabrv,references}

\end{document}